# DarSIA: An open-source Python toolbox for two-scale image processing of dynamics in porous media

Jan Martin Nordbotten[1,2], Benyamine Benali[3], Jakub Wiktor Both[1], Bergit Brattekås[3], Erlend Storvik[1,4], Martin A. Fernø[3,2]

[1] Center for Modeling of Coupled Subsurface Dynamics, Dept. of Mathematics, University of Bergen, Norway
[2] Norwegian Research Center, Postboks 22 Nygårdstangen, 5838 Bergen
[2] Department of Physics and Technology, University of Bergen, Norway
[3] Department of Computer science, Electrical engineering and Mathematical sciences, Western Norway University of Applied Sciences, Norway

**Highlights:**

- New, open-source software library for two- and multi-scale analysis of images and image series of processes in porous media.
- Seamless integration of image regularization and upscaling.
- Multi-level image alignment

## Abstract

Understanding porous media flow is inherently a multi-scale challenge, where at the core lies the aggregation of pore-level processes to a continuum, or Darcy-scale, description. This challenge is directly mirrored in image processing, where grains and interfaces may be clearly visible, yet continuous parameters are desirable to measure. Classical image processing is poorly adapted to this setting, as most techniques do not explicitly utilize the fact that the image contains explicit physical processes.

Here, we adapt classical image processing concepts to what we define as "physical images" of porous materials and processes within them. This is realized through the development of a new open-source image analysis toolbox specifically adapted to time-series of images of porous materials.

## 1. Introduction

Modern understanding of multi-phase flow and transport in porous media inherently involves discussing two fundamental scales simultaneously: On one hand, there is the relatively well-understood fluid dynamics within a complex (and in practice often unknown) pore space. On the other hand, there are the debatable effective equations at the so-called "Darcy scale", which is understood to be a scale large enough that well-sorted grain packs (such as spheres) can be treated as a homogeneous continuous material.

Using the pore scale physics to establish constitutive laws at the Darcy scale has been one of the most important and well-studied theoretical questions in the porous media community. Seminal papers in this regard go back to as far as Hubbert [1] (see also [2, 3, 4]). These theoretical developments have been supported by extensive physical and computational experimental work. From the experimental side, we emphasize techniques such as synchrotron tomography, micro-CT and PET-CT scanning, and high-resolution photography. From the computational side, we emphasize direct numerical simulation of fluid dynamics, lattice Boltzmann simulations, and network models.

However, despite the significant experimental research (both physical and computational) devoted to pore-to-core scale understanding, there is little consensus on how to actually interpret pore-scale data at the Darcy scale. A common low-order approach is to filter the data using an "averaging window", however this is a very crude regularization, which tends to both retain noise from the underlying geometry, as well as introduce significant scale-dependent effects [5, 6]. A more mathematical approach is via homogenization techniques [7], however these approaches are less suited when sharp gradients are present (such as near saturation fronts). In this contribution, we extensively discuss image regularization methods as a new way to upscale data from the pore scale to the Darcy scale.

The field of image processing has evolved to include a broad range of tools for regularizing image data, applicable to both 2D and 3D images. These tools are reflected in the development of comprehensive software packages, such as the OpenCV library [8] and the scikit-image library [9]. When open-source, these software packages serve both as a common development platform for image processing research, as well as a state-of-the-art repository of tools for the users needing image processing.

Unfortunately, most standard image processing tools are inherently single scale, do not explicitly account for physical dimensions, and do not have dedicated functionality related to the prevailing two-scale structures characterizing porous media. In this contribution, we announce the development of an open-source image processing toolbox explicitly tailored for handling two-scale images of porous media, by tailored use of openCV and scikit-image and dedicated extensions. Acknowledging this specialization, the toolbox is named "DarSIA", short for "Darcy-Scale Image Analysis".

The DarSIA toolbox is designed to be able to handle single images, comparisons of pairs of images, and time-series of images, pertaining to dynamics within both 2D and 3D porous materials. Our main ambitions and contributions are:

- Realize images as "physical images", explicitly equipped with spatial maps and physical interpretation of image data.
- Recognize the special nature of pore space and porosity in images of porous media.
- Provide pore-to-core upscaling capabilities.
- Provide multi-level image alignment functionality for direct pointwise comparison between images from different times and/or experiments.

An object-oriented Python package containing initial capabilities in this direction is being provided with the publication of this work [10], as detailed in the following sections. However, this is not a report on a final product, it is intended equally as an invitation to participation[1].

The rest of the manuscript is structured as follows: In section 2, we provide the theoretical framework and the main code components of DarSIA. In section 3, we provide a concrete application of DarSIA to experimental data. Finally, we outline in section 4 the continued vision for DarSIA.

## 2. The DarSIA toolbox

Three key requirements guide the development of DarSIA, which thereby separates it from standard image processing toolboxes. First, our images are always considered representations of physical domains. Second, the physical objects are potentially multiscale, wherein the canonical example is the combination of the pore scale and Darcy scale. Third, we are often interested in comparative analysis, either in the sense of a time series, or between repeated experiments. The three next subsections provide the tools to address each of these three goals.

### 2.1. Initializing images of physical objects

Standard image formats provide a local coordinate system (usually pixels), and attach to this coordinate system an observable function (usually color, radioactive intensity, or similar). This is a representation of a reality, which in itself can be equipped with a coordinate system in physical space (given e.g. in SI units), and providing some physical signal (such as reflection of light, radioactivity, or similar). To make this connection, it is necessary to provide transformations of both coordinates, as well as function values from the image to the physical system. This is illustrated in Figure 2.1, and we formalize this as follows:

**Definition 1:** By an $n$-dimensional *physical image* of a physical object in $m \geq n$ dimensions, we refer to the following construction:

1) An image domain $Y \subset \mathbb{R}^n$, represented by an intensity of $p$ observables $c(y): Y \to \mathbb{R}^p$.
2) A physical domain $X \subset \mathbb{R}^m$.
3) A smooth bijective mapping $\phi: Y \to X$.
4) A minimum of $q \geq 1$ *interpretations* $f: \mathbb{R}^p \to \mathbb{R}^q$.

The above construction allows us to directly refer to the physical image as $f(x) \equiv f\left(c(\phi^{-1}(x))\right)$.

*Notation.* In the remainder, the single components of vector-valued interpretations $f$ are indexed with indices $j = 0, \ldots, q-1$, i.e., the first interpretation is $f_0$. Moreover, while we in the abstract sense allow for arbitrary dimensions, in practice we expect that $2 \leq n \leq m \leq 4$, i.e. at a minimum 2D images, and at most 3D timelapse.

---

[1] Active code development on https://github.com/pmgbergen/DarSIA

When the physical image is of a porous material, we will index the interpretations from 0, and always assume that the first interpretation $f_0(x)$ is the pore-space indicator function for pore-scale images, i.e., taking values 0 and 1 for the solid and pore space, respectively. For images with noise (which is the rule, rather than the exception), the pore-space indicator function can take values within the interval $[0,1]$.

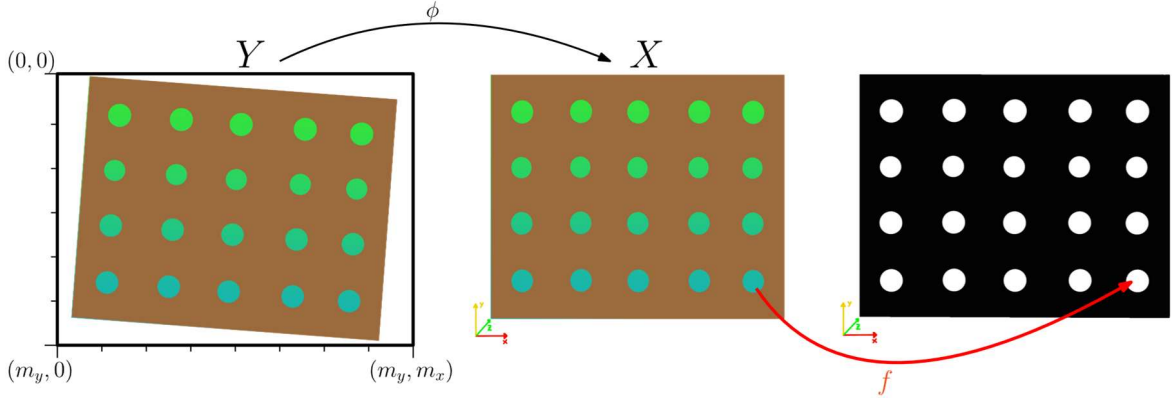

**Figure 2.1.** Illustration of Definition 1. Here, $Y$ (to the left), is the image in pixel domain, with dimension $(m_y, m_x)$, and $X$ (the middle and the right) is the physical domain. The map $\phi$ is a smooth bijection to move between the two domains. In the middle image the interpretation $f$ would simply be the identity, mapping color intensities to themselves, whereas in the rightmost image the interpretation function maps intensity values $c$ in the image $Y$ to 1 (represented by white here) if $c$ indicates a pore, and to 0 (represented by black here) otherwise.

**Remark 1**: The definition of a physical image has the colloquial interpretation that $Y$ is an image of a physical region $X$, and that every point $y$ in the image $Y$ corresponds exactly to one point in the physical object $X$. We require an interpretation of the image quantities $\mathbb{R}^p$ in terms of physical quantities $\mathbb{R}^q$. For standard trichromatic pictures, then $p = 3$ and image quantities may be the red-green-blue channel intensity values, with the interpretation $f_0(x)$ being the pore-space indicator and the interpretation $f_1(x)$ being the presence of a monochromatic tracer within a porous structure (i.e., $q = 2$).

As can be seen from Definition 1, the concept of a physical image provides additional spatial structure relative to the standard notion of an image. This additional structure is typically not encoded in standard image file formats (although we highlight formats as DICOM which contain equivalent information), and the first key functionality of DarSIA is related to equipping image files with the structure needed to be treated as physical images.

We emphasize that the generality allowed by the definition of the mapping $\phi$ is required whenever the coordinates in the image (pixels) are not simply a re-scaling of physical coordinates. This can arise due to lens distortion, or other processes associated with the image acquisition.

In conventional image analysis software, an image is typically represented as an array of $p$-valued intensities, $c \in \mathbb{R}^{m_y \times m_x \times p}$, where each element of the array $c$ is referred to as a pixel. In DarSIA, that array is complemented with information about the physical entities that the image represents. With that in mind, the initialization of a DarSIA image requires two ingredients: The image array, and information about the physical representation of the image. For two-dimensional images, it is sufficient to specify the width $w$, height $h$, and origin $o$ (position of the lower left corner of the image in the chosen coordinate system), to initialize a DarSIA image (more advanced initialization will be described later).

```
import darsia
image = darsia.Image(c_arr, width = w, height = h, origin = o)
```

**Code 2.1.** Import of the Python package DarSIA and initialization of a "physical image" in DarSIA, with an image array `c_arr` = $c$, width `w` = $w$, height `h` = $h$, and origin `o` = $o$.

Once an image is initialized, the metadata of the image is stored and can be saved together with the image array by using the built-in save method. If it is desirable to open an identical image at a later stage, the metadata and image array can be read directly at image initialization.

Furthermore, at initialization, a coordinate system for the image is created. Consequently, it includes a map $\phi$ between the image $Y$, which in the discrete setting is accessed via conventional matrix indexing, and the physical image $X$, which is represented in conventional physical Cartesian coordinates. With the above initialization the map $\phi$ constitutes a multi-dimensional scaling. This enables the user to access regions of, and positions in, the image based on physical coordinates.

```
subregion = darsia.extractROI(image, ll, ur)
```

**Code 2.2**. Extraction of a rectangular subregion from the physical image via physical coordinates. Here, `ll` is the lower left corner of the subregion and `ur` is the upper right corner.

Using the functionality to extract subregions a patch-structure is implemented in DarSIA. By specifying the number of patches that the image should be subdivided into, and how much overlap they should have, one can obtain an array of images, each one representing one patch. Each of these images can then be modified separately, before they, if desired, are glued back together to give a new representation of the full physical image.

```
patches = darsia.Patches(image, num_v, num_h)
patches[i][j] = some_modification
new_image = patches.assemble()
```

**Code 2.3.** Here, a patch object, with `num_v` · `num_h` patches, is created from the image, where `num_v` and `num_h` are the number of patches in the vertical and horizontal direction, respectively. Each individual patch is then a separate physical image and can be accessed following a matrix indexing convention. Moreover, every patch can be modified as seen fit. Here, `some_modification` is a placeholder for some image that is to replace patch `[i][j]`. Finally, the patches can be reassembled by applying the `assemble` method; overlaps are combined using a convex combination.

DarSIA allows for input array $c$ is in standard color formats. Moreover, it is seamless to switch between trichromatic or restrict to monochromatic color spaces with built-in methods, i.e., choose a particular interpretation of the physical image, relevant for further analysis.

*Advanced initialization.* In practice, images often come with systematic defects and varying

```
image.toRGB()
image.toHSV()
image.toGray()
```

**Code 2.4** Color space transformation of a given physical image, here exemplarily to the Red-Green-Blue color model, Hue-Saturation-Value color model and grayscale.

inconsistencies, with the latter in the context of time-series of images. A careful image analysis requires a unified setting for all images, and thus a range of corrections is provided by DarSIA, including corrections of systematic geometric distortions, fluctuations in intensities, image alignment, as well as non-trivial pore-space deformations, briefly elaborated on in the following.

Geometric distortions often occur due to imperfect image acquisition. Typical issues include choosing a larger frame than the region of interest, a non-planar physical asset, non-centered positioning of the camera to the object, and automated corrections performed by the camera or scanner. These create an undesired representation of the shape of the physical object in the image as illustrated in Figure 2.1. DarSIA provides tools to accurately correct for such effects using cropping, perspective transforms, stretching, and bulge correction of the image. Available reference points are vital for the accuracy. To assist fitting reference points, there exists functionality to add grids to DarSIA images, which represent the Cartesian coordinate system.

```
grid_image = image.add_grid(dx = v, dy = h)
```

**Code 2.5.** Addition of a Cartesian grid to the DarSIA image with distances `v` and `h` between vertical and horizontal grid lines, respectively.

As an example, color and illumination fluctuations in photographic images can naturally occur, e.g. if the camera is not calibrated or the scenery is illuminated under varying conditions. Such fluctuations directly affect the registration of colors and their intensities, provided by $c$, which are important for further analysis. To account for this, DarSIA can map each input array $c$ to a corrected array $\tilde{c}$, where the map is defined to *optimally* (in terms of a suitable metric) transform a reference portion of the image to pre-determined reference colors, either related to defined standard colors or a reference image, i.e. $\tilde{c}[k] - \tilde{c}_{ref}[k]$ should be as small as possible. Here, $k$ represent the indices of the reference portion of the image, and $\tilde{c}_{ref}$ denotes a set of reference colors. To aid in this, functionality from the open-source library color-science [11] is utilized.

Finally, in view of a multi-image analysis, alignment of images to some reference image is essential. DarSIA provides two different functionalities to align images, varying in their cost and their accuracy. First, using a limited set of fixed points, alignment in form of an *optimal* translation aiming at matching these fixed points can be performed. More involved, a more general deformation map can be determined, aligning the pore spaces of different images. This functionality constitutes an important analysis tool and will be explained in more detail in Section 2.4.

In sum, all mentioned corrections can be applied at initialization of physical images, providing the possibility for fine-tuning aside of semi-automatic corrections, in reference to a baseline image.

```
# Single-image correction
color_c = darsia.ColorCorrection(roi_color)
curv_c = darsia.CurvatureCorrection(config)

# Correction with respect to a physical reference image
drift_c = darsia.DriftCorrection(reference_image, roi_drift)
deform_c = darsia.DeformationCorrection(reference_image, patches)

# Advanced image initialization of a secondary physical image
secondary_image = darsia.Image(
    c_arr,
    color_correction = color_c,
    curvature_correction = curv_c,
    drift_correction = drift_c,
    deformation_correction = deform_c
)
```

**Code 2.6.** DarSIA images can also be initialized with corrections applied, also with respect to a reference image. Here, `c_arr` is the input array, `color_c` is the color correction, with knowledge on some reference region in the domain, `curv_c` is a shape correction object containing all fine-tuning parameters for crop, bulge and stretch operations (containing this time all information on the geometry including width, height, origin, cf. Code 2.1), `drift_c` is a drift correction aligning the image with a reference image using a simple translation, and `deform_c` aligns the pore space with respect to a reference image, cf. Sec. 2.4.1. Here, the term `secondary_image` is central in the context of multi-image comparisons, cf. Sec. 2.4.

**Remark 2 (Supported data types).** DarSIA currently supports various standard data types as input data, besides being able to read in images as arrays. In particular, PNG, JPG, and TIF images are supported. Functionality to read basic metadata such as a timestamp is provided, which may be used for the analysis of time-series of images. Ongoing developments aim to allow for data formats with native support for spatial meta-data (such as DICOM).

**Remark 3 (Object-oriented code design).** The object-oriented design of DarSIA enables effective caching and book-keeping, allowing for both a slim interface and increased performance for multi-image analyses.

## 2.2 Regularization of physical images and scale change

In the context of classical image processing, a regularized image $\bar{c}(y)$ refers to identifying a compromise between a measure of fidelity to the original image $c(y)$, and a penalization of noise. This is achieved by balancing two metrics, where the metric we denote as $D$ is a measure of the deviation between the images, while the metric we denote as $N$ is measuring the noise of an image. We equip the noise metric with a regularization parameter $\mu$, which is usually a linear weight providing the balance between the two metrics. In this framework, the regularized image is defined as [12] [13] [14]

$$\bar{c}(y) = \arg\inf_{u} D(c, u) + N(u; \mu_c) \tag{1}$$

While it is seldom emphasized in the image processing literature, to get resolution-independent regularization, the regularization parameter $\mu_c$ must be scaled appropriately both in terms of pixels and color scale.

The regularization parameter becomes equipped with a physical meaning when applied to physical images. We proceed by restating the definition of a regularized physical image as

$$\bar{f}(x) = \arg\inf_{u} D(f, u) + N(u; \mu_f) \qquad (2)$$

For the sake of the argument, let us for the moment consider the common choice of $D$ and $N$ leading to the Rudin-Osher-Fatemi (ROF) regularization [15], namely the weighted $L^2$ norm and $W^{1,1}$ seminorm (also known as the total variation), respectively:

$$D(f, u; \omega_f) = \tfrac{1}{2}\|f - u\|^2_{2,\omega_f} = \tfrac{1}{2}\int \omega_f (f - u)^2 \, dV \quad \text{and} \quad N(u; \mu_f) = \mu_f |\nabla u|_1 = \mu_f \int |\nabla u| dV \qquad (3)$$

For these terms to be meaningfully added together, we note that the regularization parameter $\mu_f$ must have units of $[LF]$, where $F$ is the unit of $f$ while $L$ is the unit associated with the spatial length scale of the coordinate system for $X$.

Thus, when regularizing physical images, the regularization parameter $\mu_f$ is independent of image resolution, but has the interpretation of defining a lower threshold of observation. Concretely, if $f$ represents concentration (unitless, and bounded between 0 and 1), then $\mu_f$ has units of length. Moreover, considering a physical image of a porous media, we can associate with the physical domain a length scale $\ell_{pore}$, which we interpret as a characteristic pore diameter. Now for $\mu_f \ll \ell_{pore}$, the regularized image will retain the porous structure, and will only filter out oscillations at much finer frequencies. *The result is a pore-scale image*. Conversely, for $\mu_f \gg \ell_{pore}$, the regularized image will now have filtered out information at the pore-scale, and only contain information at the Darcy scale. *The result is a Darcy-scale image*.

Based on a single pore-scale image $c(y)$ of a porous material, simple regularization therefore allows us to extract both a regularized pore-scale physical image which we denote $g(x) = \bar{f}(x; \omega_f = 1, \mu_f \ll \ell_{pore})$ and a relatively smooth Darcy-scale physical image which we denote $G(x) = \bar{f}(x; \omega_f = 1, \mu_f \gg \ell_{pore})$.

A ubiquitous feature of porous materials is that some variables may be only present in either the solid (e.g. mineral composition, material stress) or fluid phase (phase composition, saturation, fluid pressure). In some contexts (typically for conserved quantities), it is appropriate to regularize these variables directly from pore-scale to Darcy scale, in which case these quantities are already available as Darcy scale quantities in the image $G$. In particular, for all pore-scale physical images we note that the pore space is given by $g_0(x)$, and an appropriate definition of the porosity is simply the regularized pore space $G_0(x)$.

On the other hand, other quantities are typically regularized relative to their own phase (such as phase composition or pressure). To fix ideas, let us consider an interpretation $g_1(x)$ which is only meaningfully defined in the pore-space. In the solid, the interpretation is still defined (since we are looking at images), however the interpretation does not have physical meaning. As a result, the naïve Darcy-scale interpretation $G_1(x)$ is not physically relevant. To proceed, we must filter out the solid phase. The tools provided in the previous section can be combined to allow for this: Indeed, since the pore scale indicator function $g_0(x)$ takes values of 1 in the pore space and 0 in the solid, the product $g_1(x)g_0(x)$ erases all information from the solid phase. This suggests regularizing the image using a spatially dependent weight proportional to void space, i.e. $\omega_f = g_0(x)$. With this choice, we obtain a regularized pore-space image $G^p(x) = \bar{f}(x; \omega = g_0, \mu_f \gg \ell_{pore})$ and similarly a regularized solid-space image $G^s(x) = \bar{f}(x; \omega = 1 - g_0, \mu_f \gg \ell_{pore})$.

**Remark 4**: We emphasize that for any physical image $f(x)$ of a porous material at the pore scale, we obtain after regularization (at least) <u>four</u> different representations of this same physical image: A regularized pore-scale image $g(x)$, and three Darcy-scale images $G(x)$, $G^p(x)$ and $G^s(x)$, cf. Figure 2.2. The latter three images refer to upscaling of the full material, the pore space, and the solid space, respectively. This is conceptually consistent with standard averaging theories for porous media [1] [2] [3] [4] [5] [6].

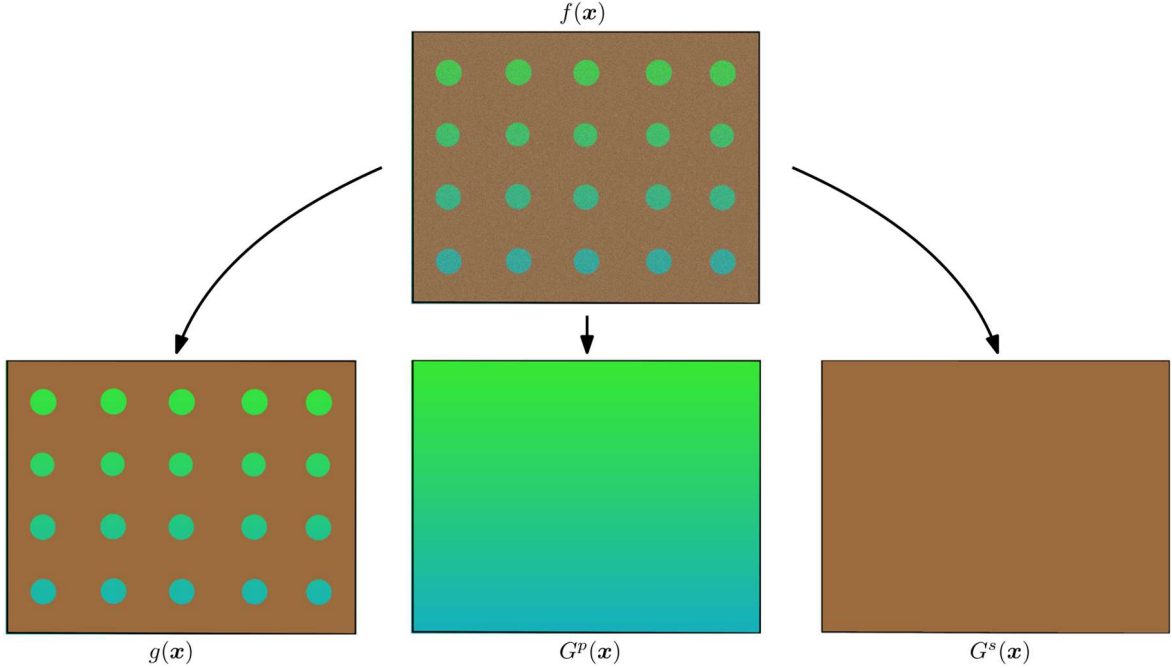

**Figure 2.2.** Illustration of the different regularizations discussed above. Here, the top image has been regularized in three different ways: $g(x)$ is the pore-scale regularization that has removed the noise that is present in $f(x)$, $G_p(x)$ is the regularized pore-space image, and $G_s(x)$ is the solid-space image.

The above discussion motivates the choice of using image regularization both as a noise removal methodology, as well as our framework for changing scales in images. In DarSIA, a total variation regularization technique with the aim of obtaining Darcy scale images is implemented.

```
regularized_image = darsia.tv_denoising(image, omega, mu, l)
```

**Code 2.7.** TV denoising using the split Bregman algorithm *[16]* with `omega`= $\omega_f$ (spatially varying), `mu`= $\mu_f$ (spatially varying) and `l` being a penalization parameter, related to the specific algorithm.

Here, an important extension of standard regularization implementations is the possibility to allow for spatially dependent metrics as in equation (3). The spatially dependent (pixel-wise defined) regularization parameter $\mu_f(x)$ is particularly useful when the aim of the regularization method is to obtain a Darcy scale image, as the regularization parameter $\mu_f$ must be of correct magnitude to remove pore-scale features from the image $\mu_f \gg \ell_{pore}$. If the image has regions of clearly different grain sizes one uniform regularization parameter might not be feasible, as too large choices will lead to overly regularized images, and even Darcy scale features will become blurred. To solve the modified total variation regularization problem, the split Bregman is implemented in DarSIA, and the noise term is decomposed to arrive at a formulation that is the anisotropic total variation denoising method, see [16], extended to spatially varying penalization.

Finally, we mention that there are several good implementations of total variation denoising for constant $\mu_c$ available in open-source Python libraries, which are suitable for obtaining $G(x)$. Scikit-image [9], in particular, contains several standard methods such as the split Bregman for isotropic and anisotropic regularization as well as Chambolle's algorithm [17]. While these implementations are more efficient than the one in DarSIA, they do not at the time of writing allow for spatially dependent regularization parameters, and therefore are not directly suitable for calculation of $G_s(x)$ and $G_p(x)$.

## 2.3 Single-image analysis tools

As pointed out in Remark 4, from a single pore-scale image, we essentially consider four different regularizations of this image, which corresponds to our best representation of spatial data. The analysis of these images will frequently be strongly case dependent (as exemplified in Section 3), and moreover, standard tools are already available for analyzing spatial data. As such, tools for analyzing single regularized images are not a main emphasis of DarSIA: Nevertheless, interfaces to useful tools provided by sci-kit image [9] are available.

One useful category of tools for single images, being highlighted here, is the geometric segmentation and image labeling, i.e., the identification and tagging of specific details in an image. For images with the structure of composites, as e.g., heterogeneous media consisting of facies, a common task is to dissect the image into the single homogeneous regions. If these are characterized by distinct intensity values, thresholding results in their detection. If, on the other hand, homogeneous regions are characterized by a range of values, shared by other regions, and merely jumps in the intensity values are present along interfaces of the homogeneous regions, a gradient-based approach instead can distinguish between the homogeneous regions (low gradient modulus) and the interfaces (high gradient modulus). Together with a watershed segmentation algorithm [18], again the homogeneous regions can be identified. DarSIA provides several gradient-based approaches varying in the degree of the user-input rewarding more control, helpful for instance for images taken in non-uniform illumination conditions.

```
labels = darsia.segment(c_arr)
```

**Code 2.8.** Gradient-based geometric segmentation and image labeling of an image, utilizing a watershed algorithm. Fine-tuning of the performance is possible through several options and parameters; here not exemplified.

## 2.4 Multi-image comparative analysis tools

Most porous media research concerns dynamics, either in the sense of transport within a phase, displacement of a phase by another, or even the hydromechanical coupling between fluids and solids. Furthermore, both physical and computational experiments are frequently repeated with slight variations between configurations, which motivates comparisons between separate experiments. As such, multi-image comparative tools are of great interest. For the sake of exposition, we will in this section consider comparative tools for two images, however, everything discussed naturally extends to any finite number of images.

### 2.4.1 Aligning pore spaces

Let $g^R(x)$ and $g^S(x)$ be two regularized pore-scale images of the same porous material, the former the "Reference" image and the latter one (or many) "Secondary" image(s). We will assume that the experimental design includes sufficient spatial reference points and that the coordinates of the images nearly align. However, in practice there will almost always be some disturbances in the geometry even for a rigid porous material, so that the pore spaces $g_0^R$ and $g_0^S$ need not be identical. The first analysis step will therefore often be to find a continuous mapping $\psi : X \to X$ such that $g_0^R(x) = g_0^S(\psi(x))$, and where we make the a priori assumption that $\psi \approx Id$, the identity transformation. Such a mapping $\psi$ is necessarily not unique (any localized deformation within the pore-space leaves $g_0^S(\psi(x))$ unaltered), and as such, we are interested in a mapping $\psi$ which in some sense is regular (or has relatively "low energy" in view of a minimization problem). Our approach to constructing such a regular mapping is through a flexible divide-and-conquer algorithm. In view of a practical algorithm, affine and perspective maps play a particular role, as for such efficient implementations exist to warp images, e.g. provided by the openCV library [8]; using such, DarSIA provides the capability to warp images with globally continuous, patch-wise defined affine maps.

*Divide-and-conquer strategy.* The main idea is to decompose the pursuit for a global mapping $\psi$, defined on $X$, into finding local translations, defined on patches of the images, and then assembling them by interpolation. The approach allows for a straight-forward multi-level extension. In the following, each component of the approach is explained in further detail.

*Partitioning.* First, the domain $X$ is decomposed into patches $p_j = 1, \ldots, P$, resulting correspondingly in subimages $g^{R,j}$ and $g^{S,j}$ of the reference and secondary image, cf. Figure 2.3 (using $I = 1$).

*Local mappings.* For each patch $p_j$, the aim is to efficiently find a low-energy mapping $\tilde{\psi}_j : p_j \to \mathbb{R}^d$ in the space of perspective mappings with domain $p_j$, here denoted by $\Psi_j$. Using a localizing metric $D_j$ (to be discussed in some more detail further below), $\tilde{\psi}_j$ is defined as minimizer

$$\tilde{\psi}_j = \arg \inf_{\psi' \in \Psi_j} D_j \left( g^{R,j}(x), g^{S,j}(\psi'(x)) \right)$$

We note that no compatibility requirement across different patches is imposed. Yet, to ensure high fidelity, the mappings $\tilde{\psi}_j$ are only trusted if they are in fact effectively just translations, resulting in situations in which patches do not get assigned a local mapping.

*Globalization step.* The local mappings $\tilde{\psi}_j, j = 1, \ldots, P$, are combined to define a global auxiliary mapping, defined on the entire domain $X$, by employing radial basis function (RBF) interpolation. A significant advantage of RBF interpolation is its mesh-free character not posing any strong requirement on the input. Thus, as input all high-fidelity pairs of the center coordinates and effective translations of the patches are used, together with additional conditions based on expert-knowledge. This can for instance be homogeneous boundary conditions in normal direction, when the porous medium is known to be fixed in certain directions. The interpolation then provides a smooth auxiliary function $\tilde{\psi} : X \to \mathbb{R}^d$, which also continuously fills-in on low-fidelity patches.

*Piecewise affine interpolation and efficient application.* By construction, the RBF interpolation $\tilde{\psi}$ satisfies $g_0^R(x) \approx g_0^S(\tilde{\psi}(x))$. Despite this being the theoretical aim, due to the generality of $\tilde{\psi}$ there exists no efficient off-the-shelf algorithm for their evaluation to large arrays, resulting in the warping of images. However, in view of the access to efficient algorithms to warp images [8], we use a now mesh-based, globally continuous, piece-wise (on patches) affine interpolation of $\tilde{\psi}$ on a Cartesian grid provided by patches. Finally, $\psi$ can be efficiently evaluated on each patch. Similarly, $\psi^{-1}$ can be approximated and evaluated.

This just described alignment procedure is implemented in DarSIA.

```
# Define analysis object based on a reference image and partitioning
deformation_analysis = darsia.DeformationAnalysis(
    reference_image,
    patches
)

# Find the global map ψ̃ to match the pore-spaces of two images
deformation_analysis(secondary_image)

# Apply piecewise affine transformation ψ aligning both images
transformed_image = deformation_analysis.apply(secondary_image)
```

**Code 2.9.** DarSIA implementation for aligning pore-spaces based on patch-wise comparison.

*Limitations and technical requirements.* The divide-and-conquer approach uses efficient feature detection and matching based on the ORB algorithm [19] combined with a RANSAC algorithm [20] to search for perspective bijection $\tilde{\psi}_j$ between the features, while discarding outlying wrongfully matched feature pairs. This poses the requirement that the patched subimages $g^{R,j}$ and $g^{S,j}$ contain sufficient resolution and size and thereby a sufficient amount of distinct features. Another natural lower bound on the patch size is given by the two images themselves and their associated deviation $\psi$; the subimages can only be compared if they contain the same features. There exist two possibilities to mitigate the need for a low-resolution search space for $\psi$. One is to use overlapping patches. The other is an iterative multilevel extension of the above idea, in principle aiming at detecting even quite general large deformations.

*Multi-level extension.* Through successive updating of the secondary images, the above divide-and-conquer algorithm can be straight-forwardly extended to an iterative multi-level algorithm.

For this, consider an $I$-multilevel partition of patches $p_{i,j}$ of $X$, $i = 1 \dots I$, $j = 1 \dots P_i$, cf. Figure 2.3.

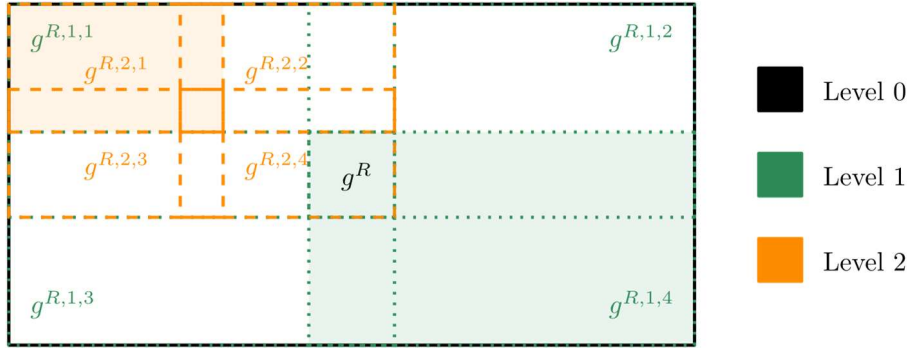

**Figure 2.3.** Multilevel overlapping partitioning $\{p_{i,j}\}_{i,j}$ of $X$ (for $I = 2$) and corresponding subimages of the reference image $g^R$ used in the 2-level divide-and-conquer algorithm, here denoted by $g^{R,i,j}$.

While iterating through the levels $i = 1 \dots I$, one keeps track of a currently best (auxiliary) global map $\tilde{\psi}^i$, starting with the natural initial guess $\tilde{\psi}^0 = Id$; a corresponding piecewise affine interpolation $\psi^i$ shall be available for $i = 0 \dots I$. Then at each level $i = 1 \dots I$, an auxiliary global map $\tilde{\psi}$ is determined using the above divide-and-conquer algorithm, now based on the partition $\{p_{i,j}\}_j$ and the pore-scale images $g_0^R$ and $g_0^S(\psi^{i-1}(x))$. The overall $i$-level approximation $\tilde{\psi}^i$ is then recursively defined as $\tilde{\psi}^i = \tilde{\psi} \circ \tilde{\psi}^{i-1}$, where with a slight abuse of notation, the composition of the two functions denotes the action of mesh-free RBF interpolation of those the composition in all interpolation points. Thus, the $i$-level iteration can be summarized in compact form, as searching for $\tilde{\psi}^i = \tilde{\psi} \circ \tilde{\psi}^{i-1}$ such that (in some sense)

$$\tilde{\psi} = \arg\inf_{\psi'} D\left(g^R(x), g^S\left(\psi'\left(\psi^{i-1}(x)\right)\right)\right) \tag{5}$$

where the piecewise affine interpolation $\psi^{i-1}$ corresponds to $\tilde{\psi}^{i-1}$ defined as the successive RBF interpolation all $i$-level low-energy mappings. Finally, we set $\tilde{\psi} = \tilde{\psi}^I$, and obtain $\psi$ as the piecewise affine interpolation. We note that due to the use of mesh-free RBF interpolation to communicate information between the different levels, there is no compatibility condition on the hierarchy of the $I$-level partition $p_{i,j}$. Yet, successively refined patches constitute a natural choice, in particular allowing to determine large deformations with high accuracy.

```
analysis.multilevel(secondary_image, multilevel_patches)
transformed_data = analysis.apply(reference_data, inverse=True)
```

**Code 2.10.** Multi-level variant for aligning pore-spaces, and warping of reference data to the pore-space of the secondary image, by applying $\psi^{-1}$.

**Remark 5 (Aligning pore-spaces as analysis tool).** For a deformable porous medium, the corrections $\psi(x)$ will give a local displacement field of the material, which can be used for calculating e.g. material strain.

**Remark 6 (Transforming static data).** Data attached to the reference image, e.g. a geometric segmentation as described in Section 2.3, can be warped through $\psi^{-1}$ and attached to the secondary image, cf. Code 2.10.

**Remark 7 (Aligning pore-spaces as correction).** Finally, $g^S \circ \psi$ maps onto the reference image $g^R$. Considering the corresponding unregularized physical images $f^R$ and $f^S$, define the corrected physical image $f^{R,S}(x) = f^S(\psi(x))$ to signify that the secondary image has been locally corrected to match the physical reference image. This correction step is in fact part of the advanced initialization of DarSIA images, cf. Code 2.6. After aligning pore-spaces, the physical images now allow for direct comparisons direct comparisons $f^{S,R}(x) - f^R(x)$. Such are central in the analysis of pore-space dynamics, as detailed in the next subsection.

### 2.4.2 Analysis of Darcy-scale pore-space dynamics

The study of pore-space dynamics lies at the heart of porous media research. Flow processes, in particular transport, have been visualized and analyzed for decades in the imaging community, with particular emphasis on the pore-scale. DarSIA follows a similar spirit enabling imaging for analyzing the same pore-space phenomena but with an emphasis on the Darcy scale. The focus of DarSIA lies in particular on transport both of passive tracers in single-phase as well as multiphase flows, as well as material deformation, and the goal is to provide an analysis quality sufficient to be treated as a measurement technology. Darcy-scale quantities of interest are continuous quantities as concentrations and saturations as well as binary data as indicators for certain phases.

To enable imaging of transport phenomena a "visual marker" is needed on the physical side, which changes its characteristics according to dominant, ongoing physical processes: moving fronts, concentration gradients, etc. A poorly chosen marker, which gives a poor signal in the image, will necessarily increase the uncertainty of the image analysis as a quantitative tool. The choice of the marker depends on the one hand on the image taking device. As examples, for PET scanners radioactive tracers are used, while in general passive dyes are suitable for photographs. On the other hand, the marker must ensure a strong signal, with large contrast to the background medium, and be sensitive to the physics allowing an at least injective mapping between signal and quantities of interest. Such a mapping requires calibration between the experimental protocol and the image analysis for optimal results. The calibration includes the important user-input of choosing a suitable interpretation of the physical image – the choice of a suitable color channel in the context of photographs for instance.

This said, let $f^R(x)$ and $f^{S,R}(x)$, as in Remark 7, be two physical images of the same porous material with their pore-spaces aligned, wherein we presume that there is an (in some sense) good visual marker present. Furthermore, let $f_1^R(x)$ and $f_1^{S,R}(x)$ be a suitable interpretation to analyze the pore-space dynamics. The central idea is to filter out the background solid by considering the difference $f_1^{S,R}(x) - f_1^R(x)$. This relative quantity has several properties we want to highlight:

- Given that the reference image coincides with an "inactive" state, e.g., an unsaturated porous medium. Then $f_1^{S,R}(x) - f_1^R(x)$ identifies the "active" pore-space, which is sufficient to use as pore-space indicator in the regularization of $f_1^{S,R}$ as described in Section 2.2, after suitable rescaling to make it unitless and or order 1.
- Separate features in the pore-space, e.g. gas bubbles, reflecting the marker just slightly differently, can be detected by analyzing the relative quantity $f_1^{S,R}(x) - f_1^R(x)$, in particular, also in cases in which the reference image does correspond to saturated conditions. Thus, the relative quantities are the vehicle to quantify saturations through effective upscaling.

*Difference between photographs and physical images.* To compare images and the specific example of photographs (as opposed to PET images), the signal intensity in the pore-space may in general be everywhere non-zero, as only the color black coincides with a vanishing signal. Instead, the signal depends on the choice of the visual marker subject to "inactive" conditions. Thus, differences of photographs are crucial, and may in practice enter already at the step of defining a physical image, cf. Def. 1, through the mapping of color to signal $f(c)$. As an example, contrasting two images under different fluid conditions (e.g. gas filled or water filled) can give access to pore-space indicators $f_0$, as the solid phase will be the part of the image that is unaltered between the two images. We emphasize that while the actual photograph (and thus image) may be highly dependent on the visual marker, the conversion of the signal to a physically relevant pore-space quantity, i.e. the definition of a physical image, should be as independent of the choice of marker as possible.

**Remark 8:** We exemplify the simplest construction of a physical interpretation from differences between images by a simple thresholding, i.e., interpretations are given as physical images with binary data

$$f_1: c^S(x) \mapsto \chi_I\big(c^S(x) - c^R(\psi^{-1}(x))\big)$$

where $\chi_I$ denotes the characteristic function of some interval $I$. Such thresholding may typically be applied to define the pore space indicator $f_0$. A more complex example is an affine conversion subject to lower and upper cut-off, for instance to define a continuous volumetric concentration like variable

$$f_2: c^S(x) \mapsto \min\big(\max(\alpha \cdot \big(c^S(x) - c^R(\psi^{-1}(x))\big) + \beta, 0), 1\big)$$

Here, $c^S(x)$ and $c^R(x)$ denote secondary and reference photographs, and we emphasize the presence of $\psi^{-1}$ to align pore-spaces. Many more conversion models are possible and depend highly on the used visual marker.

To assist in the definition of physical images requiring a reference image (e.g. a photograph), DarSIA provides functionality to define the two interpretations $f_1(x)$ and $f_2(x)$, as defined above, their Darcy-scale pore-space variants, as well as calibration tools, incl. extensions for the case of heterogeneous media.

```
# Fix reference image, and regularization and thresholding parameters
binary_concentration_analysis = darsia.BinaryConcentrationAnalysis(
    reference_image,
    regulzation_parameters,
    thresholding_patameters
)

# Determine phase location
indicator = binary_concentration_analysis(secondary_image)
```

**Code 2.11.** DarSIA implementation for extracting binary data as Darcy-scale physical image.

```
# Fix reference image, and parameters to define the Darcy-scale
concentration_analysis = darsia.ConcentrationAnalysis(
    reference_image,
    regularization_parameters
)

# Calibrate the linear model matching with volumetric injection
concentration_analysis.calibrate(
    [calibration_image_1, calibration_image_2, ...],
    injection_rate = val
)

# Determine physical image from a secondary image
concentration = concentration_analysis(secondary_image)
```

**Code 2.12.** DarSIA implementation for extracting concentration-type data as Darcy-scale physical image. The conversion is calibrated based on a set of images which is supposed to coincide with a volumetric injection for some rate. The calibrated converter finally determines the concentration corresponding to an arbitrary, secondary image.

## 2.5 Visualization tools

Visualization often complements quantitative analysis, in particular in the case of image data. Therefore DarSIA also provides some functionality to visualize and post-process data determined in Section 2.4. The use of many of the following tools will be demonstrated with examples in Section 3.

We highlight the visualization of the pore-space alignment map $\psi$, cf. Section 2.4.1, which displays displacement using a glyph plot.

```
deformation_analysis.plot()
```

**Code 2.13**. Show glyph plot of the result of the deformation analysis, cf. Code 2.9.

Furthermore, provided related binary data, determined as in Section 2.4.2, associated to multiple physical images, e.g. advancing in time or from different experimental runs. Visualization of a comparison of the binary data, detecting unique appearances and overlaps between an arbitrary number of segmented images, and thereby allowing for comparing different experiments. Moreover, functionality is included to compute the fractions of each color (and thereby also the different instances of overlap) that appears in the image. These can also be weighted arbitrarily.

```
compare_segmentation_comparison = darsia.SegmentationComparison(n)
comparison_image = segmentation_comparison(seg_1, seg_2, …, seg_n)
color_fracs, *_ =
segmentation_comparison.color_fractions(comparison_image)
```

**Code 2.14.** Illustration of a set of binary data (`n` many) associated to multiple physical images. In addition, ratio of unique and overlapping segments is quantified, useful for instance to study physical variability.

Finally, in the spirit of particle tracking, an unsupervised detection of the finger tips at propagating fronts can be performed, allowing for their tracking in space and time. For this, consecutively, finger tips at two distinct times are detected and associated with each other. Aside of the possibility to study finger lengths, onset time etc., the trajectories of finger tips in space-time can be visualized.

```
contour_analysis = darsia.ContourAnalysis([seg_1, seg_2, …, seg_n])
contour_analysis.plot_finger_trajectories()
```

**Code 2.15.** Visualization of the trajectories of finger tips, traveling in space-time.

# 3. Application to 2D images of porous media

DarSIA is developed with the aim of being a general tool for analyzing experimental pore-scale data, both in 2D and in 3D. Nevertheless, the impetus for the initial development comes from the high resolution photographic data available through the FluidFlower concept and experimental program [21] [22] [23]. In this section, we therefore demonstrate in the following how DarSIA can be used to analyze two-dimensional images taken of porous media experiments conducted in the FluidFlower rigs. It should, however, be noted that the implementation of DarSIA aims at being general and other images, or experimental setups, can be treated in a similar way

The FluidFlower concept is a suite of experimental rigs at meter scale, allowing for constructing three-dimensional porous media with relatively (in some sense negligible) shallow depth, and equipped with a glass front panel, allowing to view one of the major sides of the porous medium. Hence, an essentially two-dimensional porous medium is provided. Filled with unconsolidated sand, arranged in layers, consisting of homogeneous regions, relatively complex media can be constructed, including fault-structures, and sealing caprock-like layers. Finally, various fluids can be injected as water, tracers, and $CO_2$. By using pH-sensitive dyes the fluid flow can be visually perceived. Images are acquired to monitor experiments. As such the FluidFlower rigs allow for investigating various research questions in the field of multi-phase, multi-component flows in unconsolidated porous media, while a tool as DarSIA allows for quantitative research. And, indeed, DarSIA in combination with the FluidFlower rigs has been successfully used in studying tracer flow [24] as well as $CO_2$ storage experiments [22] [23].

In the following, we present some of the workflows used to analyze the images of the aforementioned experiments, in particular those related to the International FluidFlower Benchmark study [25] [26] [27]. Key points include concepts introduced in Section 2 as a correct initialization, detection of facies, aligning pore-spaces, and investigating the $CO_2$ distribution, exemplified for two configurations (5 and 24 hours after injection start). We consider these workflows as natural when dealing with real data, and expect them to be applicable as a prototype also for other physical experiments. Indeed, the same approach has been utilized to extract continuous concentration data from photographs [24], including a model calibration as detailed in section 3.6.

In the benchmark, the largest rig within the FluidFlower family has been used, with a width of 2.8 m, a height of 1.5 m, and a varying depth of 18 – 28 mm. While being locally flat, the asset is curved, which results in strong nonlinear projections onto the two-dimensional canvas when captured with a photo camera. With a high-resolution camera (35.5 MP with relatively good signal-to-noise ratio), the grains of the coarsest sands are clearly resolved, although the finest sands are not. To enable controlled color and illumination corrections, a standardized ClassicColorChecker is attached to the rig, which DarSIA is able to utilize. Finally, the formation is saturated with water and a pH-indicator, allowing to visually distinguish between water, $CO_2$-saturated water and gaseous $CO_2$, cf. Fig 3.1.

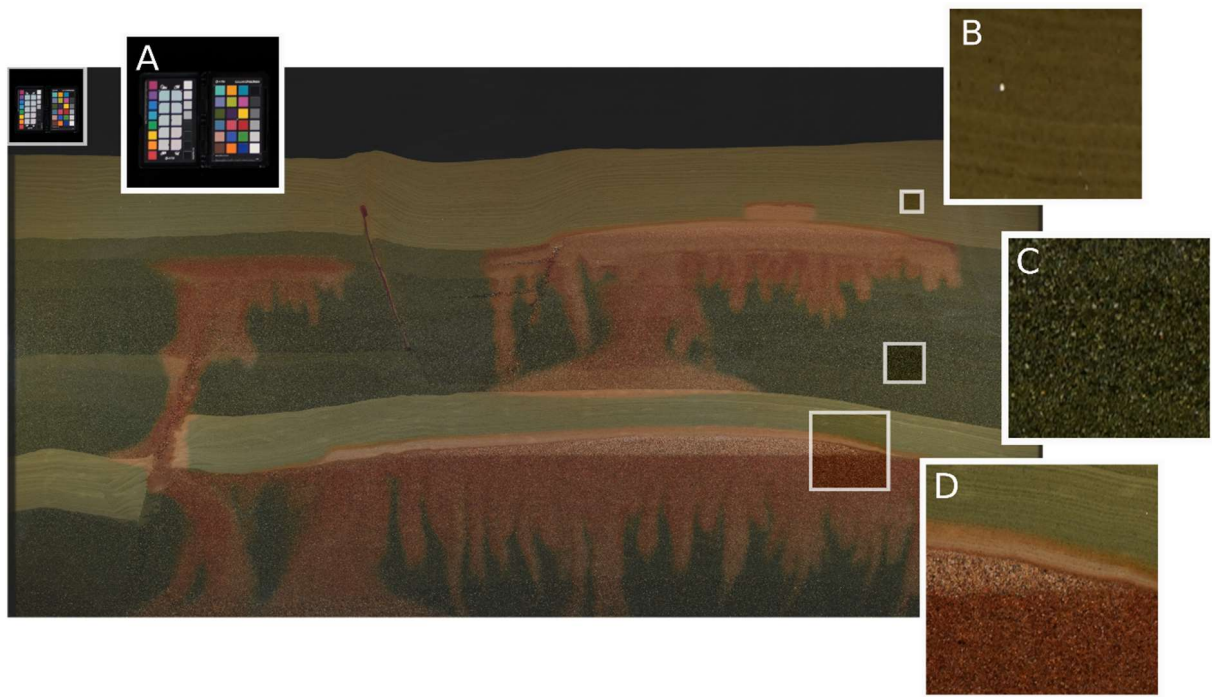

**Figure 3.1**. Illustration of input to image analysis. $CO_2$ injection in FluidFlower benchmark geometry from International Benchmark study, displayed with emphasis on relevant properties for the image analysis: (A) standardized color checker for controlled color correction; (B) bright, fine grained sand captured with relative low-resolution – single grains with shining effect; (C) darker, coarser sand captured with relative high-resolution, similar to remaining coarser sands; (D) pH-indicator reacting to presence of $CO_2$ and allowing for distinguishing between water, dissolved $CO_2$ and gaseous $CO_2$.

## 3.1 Initializing physical images

The three-dimensional physical asset has been designed as a curved entity [21], yet locally it has a natural two-dimensional character. Thus, in terms of Section 2.1, a bijection $\phi: Y \to X$ is required which is more complex than just a linear rescaling, since $Y \subset \mathbb{R}^2$ while $X \subset \mathbb{R}^3$. Using polynomial bulge and stretch corrections in addition to applying a perspective transform to map the corners of the medium onto a rectangular domain with the right physical dimensions, the seemingly non-rectangular reservoir can be distorted to an actual rectangle. Since this is not a trivial task, and the uncertainty for having sufficiently corrected the domain is relatively large without having reference points inside the domain, calibration photographs of the FluidFlower have been taken with a laser grid with uniform spacing of $10$ cm $\pm$ $1$ mm projected onto the surface. Together with the functionality in DarSIA to plot images with grids on top, the correction map $\phi$ can be determined with sufficient accuracy, thus allowing to define a Cartesian coordinate system and a corresponding, meaningful metric, see Figure 3.2.

## 3.2 Detecting facies

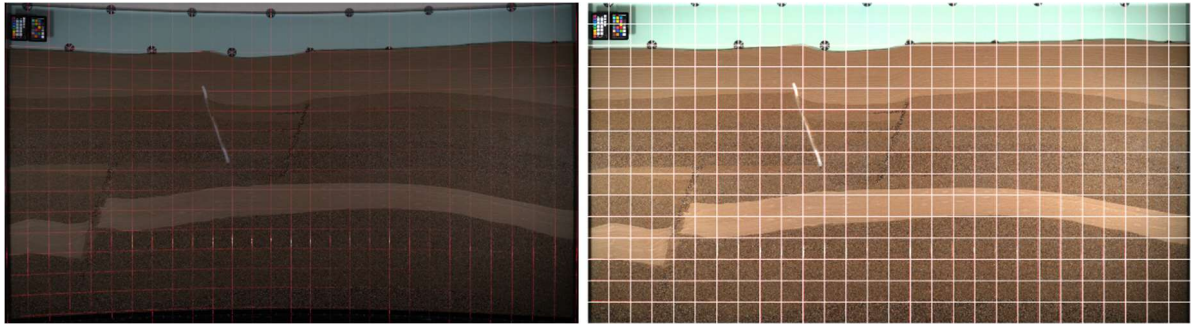

**Figure 3.2.** *Left:* Unprocessed image of FluidFlower with laser grid projected on top, distorted due to the curved shape of the FluidFlower; in addition, suboptimal illumination. *Right:* Distortion and color corrected image, with artificial grid on top to assist the fine-tuning and examine the correction result.

The FluidFlower geometry is designed to be a layered geometry built from six facies of internally homogeneous sands, where the different facies are characterized by their grain size distributions. These have in general slightly different colors and brightness and thus respond differently to the pH-sensitive dye when in contact with $CO_2$. Thus, having basic color arithmetic in mind, differences of interpretations in a multi-image analysis, as described in Section 2.4.2, will require separate analysis on each facies. Furthermore, due to non-uniform illumination, each homogeneous region has to be treated separately, instead of aiming at the same treatment for each facies.

A gradient-based geometric segmentation is used in DarSIA to detect some of the layers. Note, as discussed in the caption of Figure 3.3, the differences between the different layers in the middle part of the image are visually not very distinct. Nevertheless, the image analysis produces a coherent result, correctly identifying the main structures.

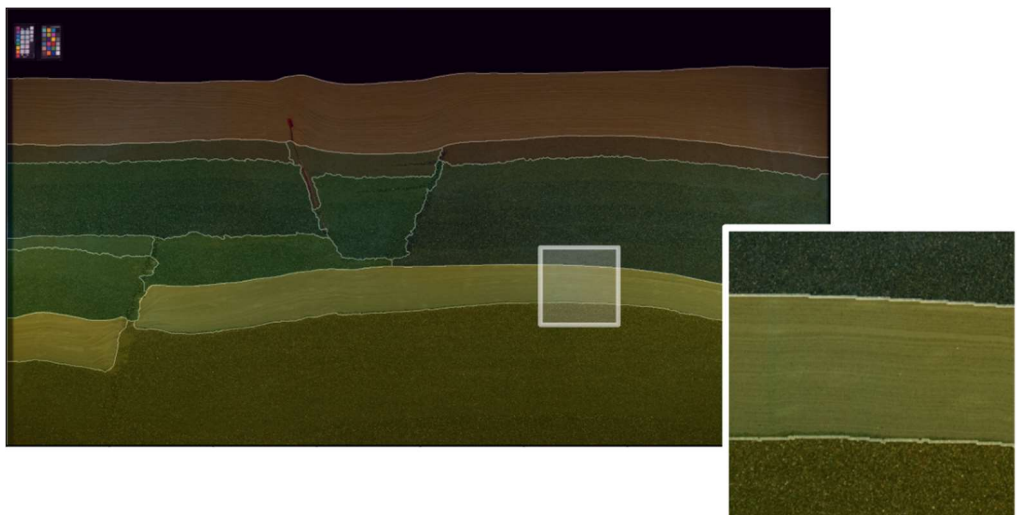

**Figure 3.3.** Segmented and labeled FluidFlower benchmark geometry. Each labeled region is colored differently and interfaces are highlighted with white lines. Several layers in the mid strip of the geometry are not segmented in full detail. Here, a merely weak intensity gradient together with the need to regularize the image to effectively perform the segmentation in the intensity puts challenges to the watershed algorithm. Zoom-in demonstrates the slight inaccuracy of detected (lower) interfaces due to the use of regularization.

Our human visual perception suggests that an accurate dissection of the medium should be possible. However, due to the minute heterogeneities between sand grains, together with the small differences between the grain size distributions between the different facies, accurate identification of sand structures is even not trivial for humans. As a retrospective comment, the identification of sand layers could be facilitated if the experiment utilized sands with greater visual contrast.

### 3.3 Aligning pore spaces

As the FluidFlower geometry is constructed from unconsolidated sands, once poured into the rig, the sand continues to settle when subjected to stresses from fluid flow. Such sand settling is undesired when wanting to study physical variability of multi-phase processes [22], but on the other hand, is an interesting observation regarding the flow of a saturated granular media. Thus, it can be of interest to precisely quantify the spatial map of sand settling through time. The multi-level feature-detection-based deformation analysis, cf. Section 2.4.1, allows for such spatial quantifications, as shown in Figure 3.4.

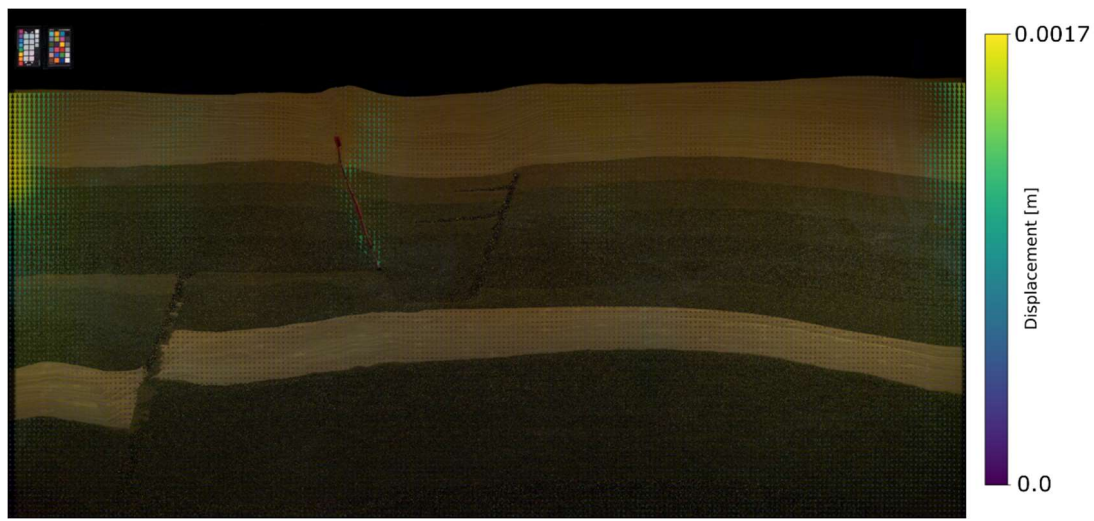

**Figure 3.4.** Effective deformation required to align pore-spaces of the configurations of two different baseline images, corresponding to the injection start of the first two experimental runs of the FluidFlower benchmark. DarSIA is able to detect local displacements of the order of 1.7 mm (and smaller), which corresponds to approximately 6 pixels. The glyph plot is obtained through the visualization functionality displayed in Code 2.13.

In order to find a deformation mapping which detects relative displacements as small as just a few pixels, the partitioning of the image has to be relatively fine. However, as explained in Section 2.4.1, finding local deformation mappings may be challenging if the resolution is not sufficient, in particular when using a fine partitioning. This effect can be observed in the analysis of the FluidFlower dataset. For the finest sand (the brightest in the images), single sand grains cannot in general be detected. Indeed, as an unintended consequence of the care with which the sands were prepared, the result is that for small patches within the finest sands, sufficient identifiable features for calculating a local deformation map may not be available. This exemplifies a situation where a multi-level deformation analysis is of great utility, as larger deformation patches will still contain identifiable features (if not within the finest sand, then at least boundaries between sand layers), cf. Figure 3.5.

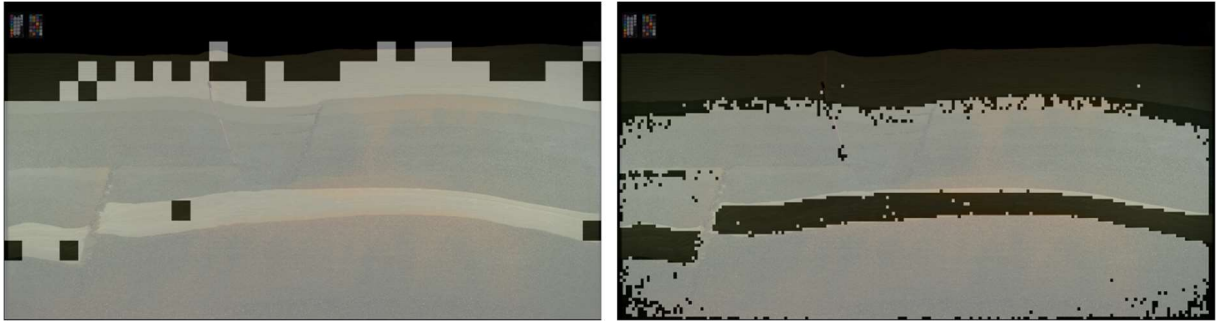

**Figure 3.5.** Illustration (debugging feature of DarSIA's deformation analysis tool) of success to find an effective translation connected associated each patch; a light patch indicates success, while a dark one indicates failure. Left: Success for coarse level analysis (32 × 16 patches). Right: Success for the subsequent fine level analysis (200 × 100 patches). Note the systematic failure for the finest/brightest sand type with lowest relative resolution per grain size.

## 3.4 Phase segmentation

Given a time-lapsed series of fine-scale images of an entire experimental run, the extraction of Darcy-scale pore-space quantities has been of large interest in the benchmark. As detailed in [22], the chosen pH-indicator has not allowed for extraction of continuous concentration and saturation profiles. Yet, instead, image analysis and the introduction of binary phase indicators as physical images, cf. Section 2.4.2, is possible. Based on the color and intensity response of the pH-indicator when in contact with $CO_2$, cf. Fig 3.1, the image can be segmented into regions identifying the three different phases: clean water, water with elevated $CO_2$-concentration, and $CO_2$ gas.

The chosen procedure is the following. First, all $CO_2$ (both in aqueous and gaseous phase) is detected, separating it from the water. Second, regions with gaseous $CO_2$ are identified as part of the overall $CO_2$. Both approaches are based on thresholding, as described in Section 2.4.2, thus from the perspective of the analyst, require choices on a suitable monochromatic signal and threshold parameters. For the latter, the knowledge on the heterogeneous structure of the medium, cf. Section 3.2, allows choosing tailored parameters for each detected layer. Here, we consider two ways of choosing the parameters: a semi-automatic dynamic and manually tuned thresholding.

### Reference image

As discussed in Section 2.4.2, the trichromatic photographs have to be related to a reference image, in order to allow conversion to binary data. Here the reference image is chosen to be the geometry, solely saturated with water. Several such images have been taken prior to the $CO_2$ injection. Despite no physical variation, differences of these reference images are not close to 0, cf. Figure 3.6. Due to illumination fluctuations in the lab, wrong "inactive" signals are detected when choosing one of the photographs as a single reference image. To mitigate this issue, essentially a pointwise maximum of all reference images is chosen to be the final reference image. This procedure in some sense serves as cleaning of the signal from systematic noise.

### Choice of a monochromatic color space

To dissect images and distinguish between water and $CO_2$ as well as gaseous $CO_2$, a suitable monochromatic interpretation of the image differences, cf. Sec 2.4.2, needs to be chosen. Detecting any $CO_2$ is identical with detecting any changes with respect to the reference image. In trichromatic (RGB) image comparisons, the solid-space should produce zero values (black in terms of color space), while any non-zero contribution can be associated to activity, thus, some $CO_2$.

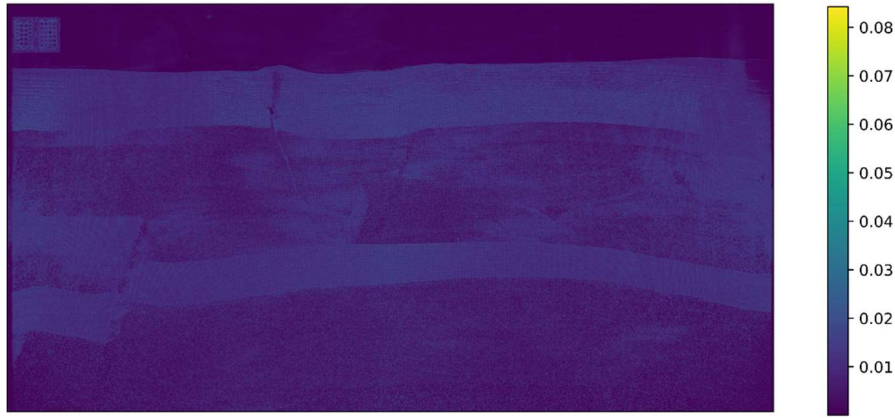

**Figure 3.6.** The pointwise maximum of 10 image differences (in the blue color channel later used in the segmentation of $CO_2$ gas), comparing secondary "reference" and a "true" reference image. The scale is in fractions of unity. Thus, all are seemingly images of the same configuration, acquired over the course of a few minutes. Reflections from both the color checker and the reservoir suggest systematic illumination variations. Moreover, each sand type may reflect light differently. Overall, in this dataset, undesired variations in the images of up to 2-3% of the RGB scale can be expected.

Using the CMYK (Cyan-Magenta-Yellow-Key) color space and specifically the Key-channel (corresponding to black), allows for a suitable monochromatic interpretation and desired segmentation based on thresholding, cf. Figure 3.7. In fact this choice appears to be agnostic to the various pH-indicators considered, which simplifies the selection when considering various pH-indicators with very different color spectra [23]. Similarly, a dedicated monochromatic color space can be selected to detect gaseous $CO_2$.

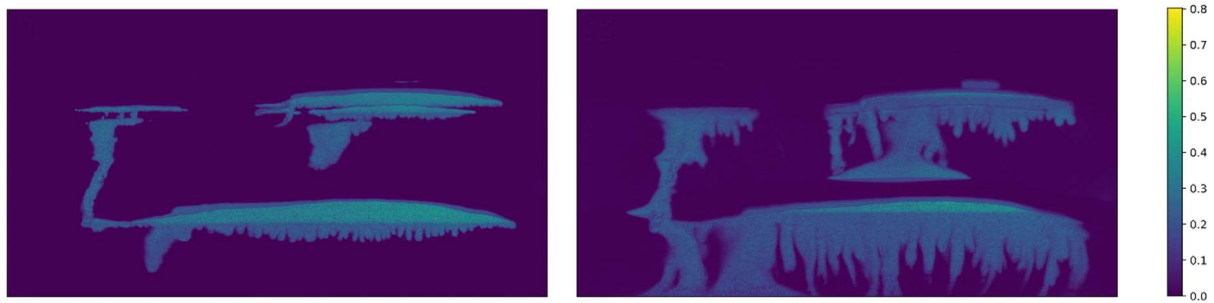

**Figure 3.7.** The negative of the key-channel of the signal difference for the FluidFlower benchmark dataset. *Left*: At the end of the $CO_2$ injection (5 hours after injection start). *Right:* 24 hours after injection start. The $CO_2$ plumes have distant signals from the water regions (almost 0). Also the gas regions seem to be characterized by a significant range of values. Yet, for the left image, one can also see, that similar intensity values are locally reaches in aqueous regions.

In an optimal setting, a monochromatic representation of the image differences would not require any thresholding to identify the desired phase, but would be a (potentially non-linear) transformation of the actual continuous concentration. However, due to limitations of the pH indicators (small activation ranges and precipitation), this is not possible for this dataset.

### Calibration of threshold parameters

After choosing a suitable monochromatic interpretation, suitable thresholding parameters have to be selected; DarSIA provides extensions of Code 2.11 and 2.12 to heterogeneous media, allowing for including the geometric segmentation from Section 3.2 as input. This is part of a calibration process. Ideally, for each sand type, a controlled sensitivity study should allow for identifying proper values, applicable for any constructed geometry. In the current context, images acquired as part of the experiments themselves were used for the calibration. The lack of controlled reference images lowers the level of automation possible, and requires a greater input of human expert knowledge and visual examination with associated fine-tuning. Using the latter calibration routine, which follows a standard training-validation-testing paradigm, there is no guarantee for robustness.

In order to aid such calibration processes, DarSIA includes a range of dynamic, unsupervised histogram-based thresholding schemes, which are close to the well-known Otsu thresholding method [28], but are tailored to scenarios in which either no "foreground" or "background" exists which should be separated, cf. Figure 3.8. The dynamic thresholding can be used in at least two ways. One can use the dynamic threshold parameters as initial guesses for further tuning. Or, if less accurate results are sufficient, for instance in a fast-prototyping environment [23], the dynamic thresholding can be even blindly used in unsupervised fashion.

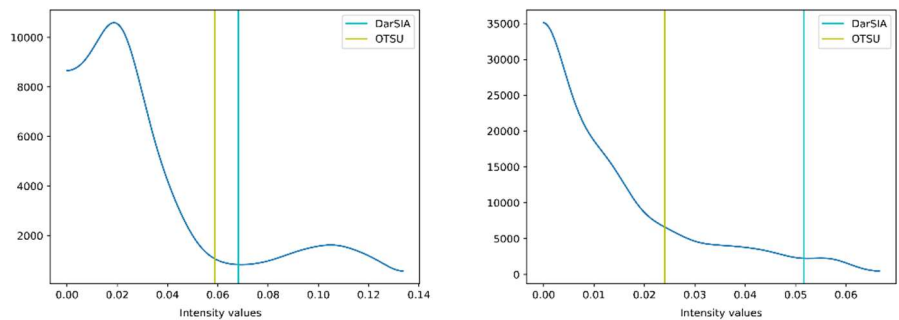

**Figure 3.8.** Histogram analysis, inspecting a single detected homogeneous region, cf. Sec. 3.2, for finding a suitable threshold parameter for detecting gaseous $CO_2$. Comparison between standard Otsu thresholding and a more careful choice provided in DarSIA. *Left:* 5 hours after injection start. Two distinct intensity peaks can be identified, representing the background (low values) and foreground (high values). Some value in between seems in principle a good candidate for constant thresholding. Both Otsu and DarSIA provide relatively similar values. *Right:* 24 hours after injection start, illustrating a typical situation in which a homogeneous region is subject to signal without clear intensity peaks. Otsu and DarSIA provide two different values, where DarSIA remains close the value predicted for 5 hours.

In Fig 3.9, a comparison of the static and dynamic heterogeneous thresholding techniques is presented. Here, the static one is calibrated to work robustly for the entire FluidFlower benchmark dataset, whereas the dynamic one is adapting to each image. Overall, both algorithms capture the main aspects of the experimental data. However, when looking at the details of the segmentation, the two algorithms have different strengths and weaknesses in various parts of the domain. We emphasize again that the ultimate result depends not only on the image analysis algorithms, but also on the chosen visual markers in the experiment.

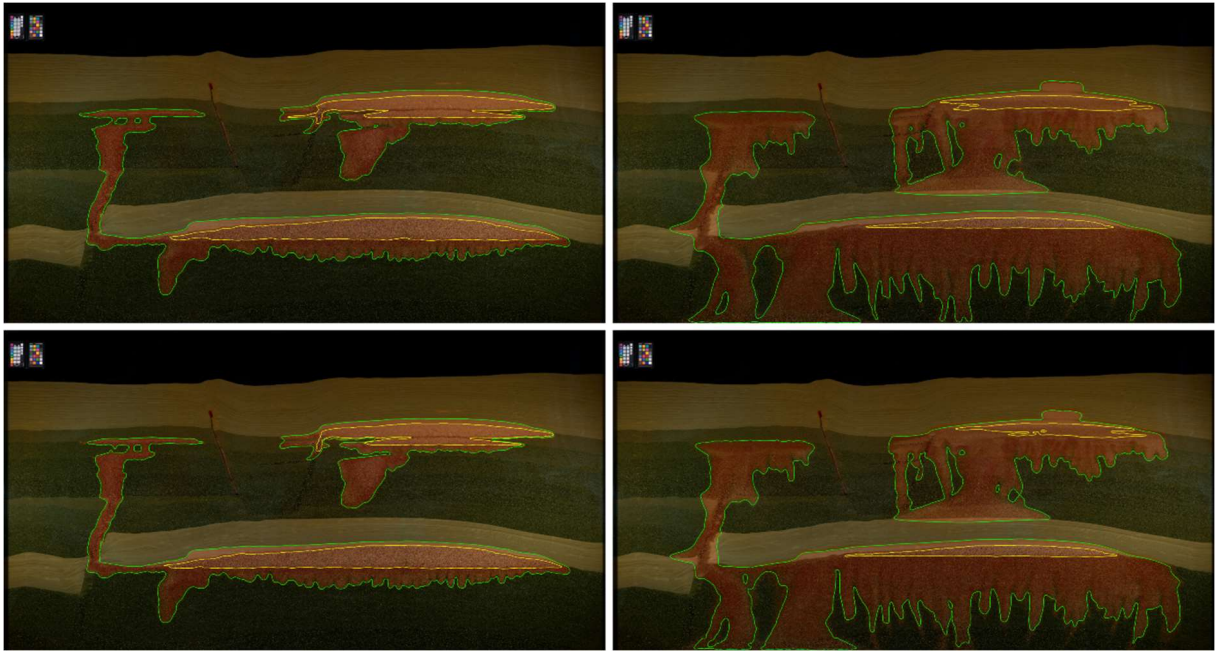

**Figure 3.9.** Phase segmentation for the $CO_2$ injection in the FluidFlower benchmark geometry. Gaseous $CO_2$ and $CO_2$-saturated water are indicated by yellow and green contours, respectively. *Top:* Static thresholding. *Bottom*: Dynamic thresholding. *Left:* 5 hours after injection start. *Right:* 24 hours after injection start. Discussion in the text.

Any dynamic geometric segmentation, including the above phase segmentation, provides possibilities for further analysis. Using visual means, the sparse nature of segmentations in terms of intensity allows for direct comparison of different configurations by simple overlapping on a single canvas, cf. Code 2.14 and Fig. 3.10.

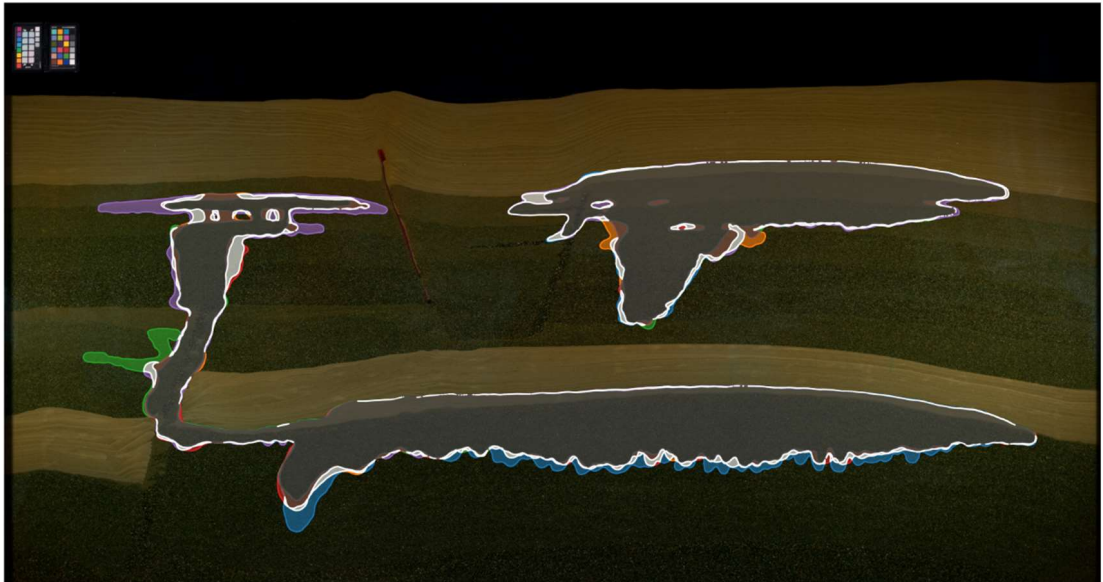

**Figure 3.10.** Visual comparison of all (five) experimental runs of the International FluidFlower Benchmark study, 5 hours after injection start. Each distinct color (besides grayscale) identify unique location of $CO_2$ in the geometry, while the gray values depict overlapping regions. Dark gray addresses all runs, while light gray includes only a handful. Visual inspection confirms that the observed variability is physical, and not an artefact of the segmentation algorithms. For further discussion on the physical variability of the $CO_2$ experiments conducted in the FluidFlower see *[22]*.

## 3.5 Density-driven convective mixing

Binary data localizing fluid phases hold information on propagating fronts. In multi-component flows, the development of density-driven fingers (signaling the onset of density-driven convective mixing), which can also be understood as unstable perturbations of a diffusive front, is a topic of interest, in particular in the context of $CO_2$ storage (see e.g. [29, 30, 31]). DarSIA provides functionality to determine extremalities of boundaries of binary data. Thus, finger tips can be effectively identified. Given time-lapsed images, the position can be furthermore tracked in space-time, and plotted, cf. Code 2.15.

The FluidFlower benchmark experiments also show such density driven fingering of dissolved $CO_2$ within the water phase. Fig 3.11 illustrates the possibility of identifying finger tips, here restricted to a certain region of interest (box C), relevant for the International Benchmark study [25], while the corresponding trajectories of the finger tips are illustrated in Fig 3.12. Besides plotting, their space-time locations can be used further to analyze various properties such as the onset of unstable finger growth, characteristic wavelength and velocity.

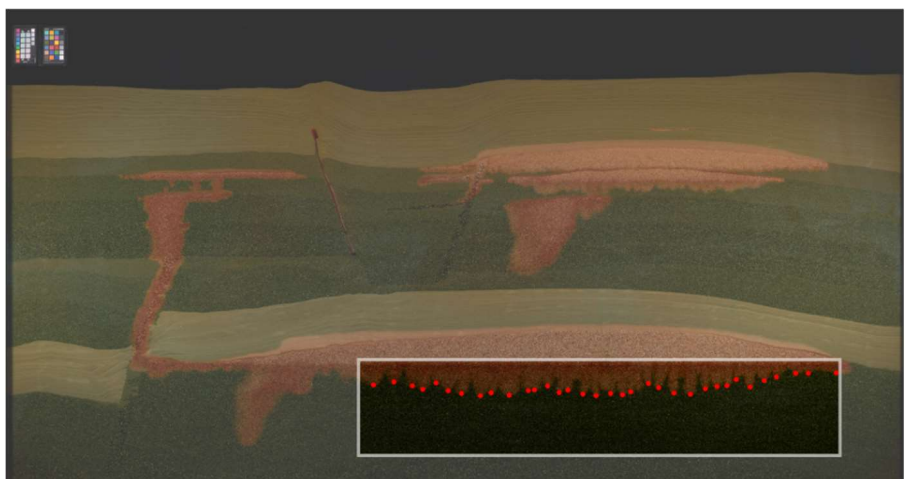

**Figure 3.11.** Finger tips 5 hours after injection start, in a region of interest (box C).

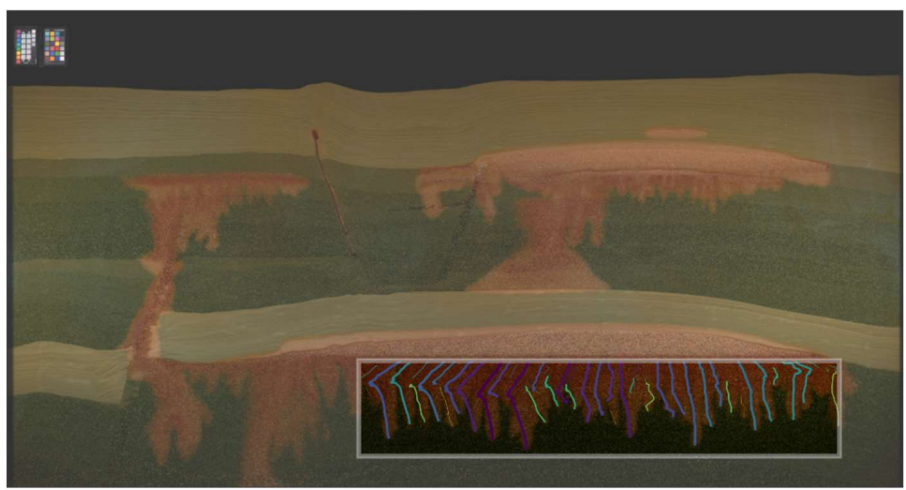

**Figure 3.12.** Trajectories of fingers, travelling through box C. The thickness of the paths relates to the relative length (compared to the longest path), for visualization purposes only to highlight dominant trajectories. Note the zig-zagging which occurs due to circulation in the formation

## 3.6 Darcy-scale tracer concentration maps

Using an analogous procedure as described in Section 3.4, DarSIA can be used to extract continuous data from 2D images of porous media. This is demonstrated in the following based on a tracer experiment in a medium-sized FluidFlower rig, cf. Figure 3.13, with similar properties as the large rig, cf. Figure 3.1. The experiment considered here [24], follows three stages: First, a monochromatic tracer is injected with constant injection rate into a homogeneous medium; second, the injection rate is increased by a factor of two; finally the injection is stopped, cf. Figure 3.14.

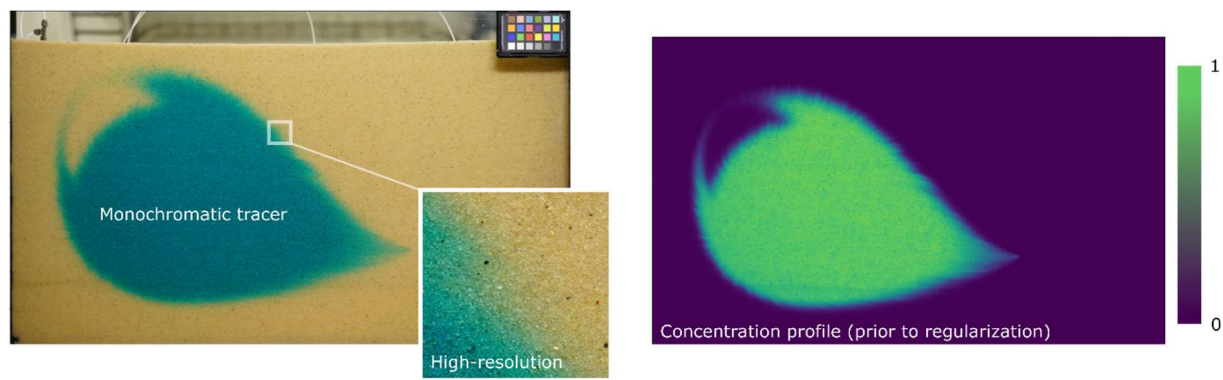

**Figure 3.13.** Tracer experiment in homogeneous medium, conducted in a medium-size FluidFlower rig. *Left:* Photograph of experiment with zoom-in on transition zone. *Right:* Converted pore-space interpretation of tracer concentration, obtained through analogous procedure as in Section 3.4, but for continuous data.

In order to extract physically meaningful data from the corresponding photographs, the procedure described in Section 2.4.2 is followed. Using image comparisons with respect to a reference image, the pore-space can be isolated, allowing for examining the tracer concentration. To correlate a monochromatic signal intensity (here the grayscale) to actual tracer concentration, a linear conversion is chosen, requiring suitable scaling parameters. These can be calibrated based on a set of calibration images, resulting in actual concentration profiles, cf. Figure 3.13. In Figure 3.14, the calibration and its validation for the described experiment is illustrated, showing its satisfactorily performance.

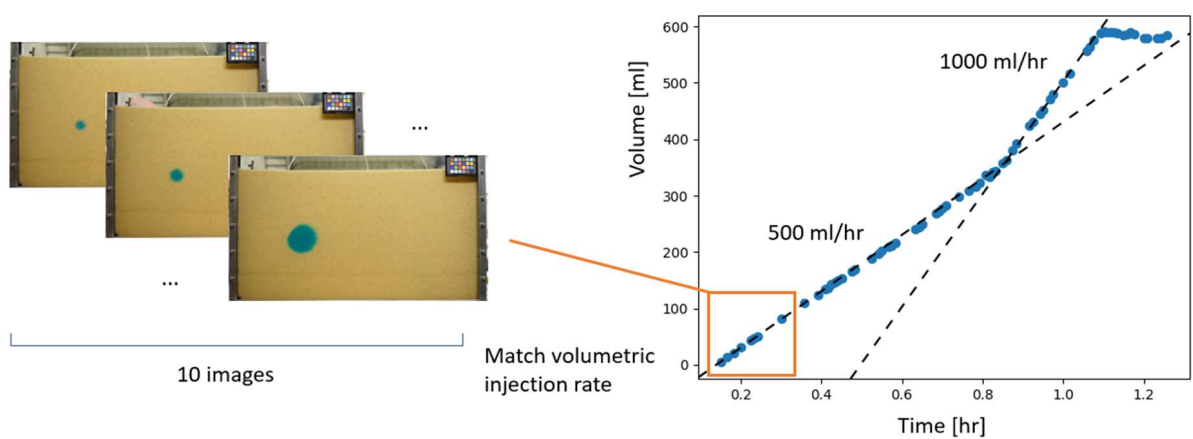

**Figure 3.14.** Illustration of calibration routine, as indicated in Code 2.12. Calibration is based on matching the effective volumetric injection rate interpreted from 10 images with the known injection rate. The calibration is validated by comparison with images, corresponding to later times in the experiment, and known injected volume, satisfactorily matching all three stages of the experiment (injection with 500 ml/hr; 1000 ml/hr; injection stop).

# 4. Conclusions and future developments of DarSIA

Our vision is the use of DarSIA as quantitative measuring instrument in porous media laboratories – both experimental and computational. Physical images, with attached physical coordinates and interpretations, allow for both software and experts to analyze images taken of static and dynamic pore-space processes. We identify that image regularization is not only a noise removal tool, but indeed provides a seamless framework for pore-scale to Darcy-scale upscaling and analysis. Together with careful choices of interpretations of physical images and sufficient calibration we show by application to the FluidFlower dataset that this allows for quantitative analyses of real-world data.

An important requirement for DarSIA to be applicable to laboratory data is a close interplay between experimental design and image analysis. We identify several factors that have to be treated with care. We emphasize in particular the necessity (and possibility) of aligning images based on well-defined reference points, as well as an appropriate choice of the visual marker for any dynamical process. These concepts are essential when direct pore-scale comparisons are desired. Concretely, if small deformations (or pore-scale processes) shall be studied using the algorithms described in Section 2.4.1 (or 2.4.2), the error in aligning the coordinate systems of two images has to be significantly smaller than the characteristic displacement (or characteristic pore diameter). Therefore, markers for image calibration has to be considered as part of the experimental design, aiming at determining conversion models and estimate all required tuning parameters. Finally, the concepts behind DarSIA are not restricted to two-dimensional images, and an extension to three-dimensional images and support to standard data types (e.g. DICOM) in the field of PET/CT-scanning is ongoing.

Recognizing the impact of careful experimental design tailored for the needs of image analysis, we envision further development of the code basis of DarSIA. Future efforts will be put in a user-friendly interface not only allowing experts to analyze images, but to enable its use ready from the start of an experiment – in its design phase. For this, we envision diagnostic tools to assess usability of visual media. For instance, the selection of a color space could be viewed as a constrained optimization problem, when searching for the optimal hue-saturation combination.

On the side of the analyst, we see a need for extended user-friendliness in terms of allowing for direct interaction with the post-processed results. A graphical user interface allowing to select reference points entering the tuning of distortion corrections or regions of interest entering definitions of color corrections would make DarSIA significantly more accessible. Having quantitative research and physical variability in mind, an extension of the suite of suitable metrics for comparing different Darcy-scale variants of physical images, we are aiming for more sophisticated distances than using a basic $L^2$-mindset for binary data. Concepts from optimal transport, as the Wasserstein distance, also used widely in other areas of image analysis, could allow for meaningful comparisons of continuous data especially for transport-dominated processes (see e.g [32]).

Complementing the above visions, further development of DarSIA also has to include improvements of the computational performance of the algorithms. The current version of DarSIA makes already tailored use of effective and scalable patching of images based on slim data structures. However, when entering the third spatial dimension, and hopefully also the fourth (time-lapse 3D data), there are clear theoretical advantages to apply regularization directly in 4D. Numerical algorithms for 4D image regularization are in no way standard (for a recent discussion, see e.g. [33]), and their development has to be both robust and run time optimized to allow for reasonably high resolution in space and time. This challenge is underscored by the observation that even for the two-dimensional images discussed in Section 3, one can make a strong case for the need of tailored and fast Darcy-scale image analysis algorithms.

A final remark pertains to automation. The complexities of imaging and porous media research imply that human-supervised use of the DarSIA components will be necessary for most research applications. On the other hand, there is still a drive in several applications towards real-time and unsupervised analysis, of laboratory data, such as in the realm of digital twin technology [24]. The desire for unsupervised applicability of image analysis further supports the argument for robustness, efficiency, and reliability of the underlying algorithms.

In closing, we reiterate our invitation from the introduction, and encourage interested parties to join the future developments of DarSIA.

## Funding

The work of JWB is funded in part by the UoB Akademia-project «FracFlow» and the Wintershall DEA-funded project «PoroTwin». BB (Benali) is funded from RCN project no. 324688.  BB (Brattekås) is funded in part by the UoB Akademia- project «FracFlow»  and RCN project no. 324818.